\documentclass[pra,twocolumn,notitlepage,showpacs]{revtex4-1}
\usepackage{graphicx}
\usepackage{amsmath}
\usepackage{amsfonts}
\usepackage{amssymb}
\usepackage{epsfig}

\begin{document}

\title{All-cavity electromagnetically induced transparency and optical switching:\\
semiclassical theory}

\author{Aur\'{e}lien Dantan, Magnus Albert and Michael Drewsen}

\affiliation{QUANTOP, Danish National Research Foundation Center for Quantum Optics, Department of
Physics and Astronomy, University of Aarhus, DK-8000 \AA rhus C., Denmark}

\begin{abstract}
The transmission of a probe field experiencing electromagnetically induced transparency and optical
switching in an atomic medium enclosed in an optical cavity is investigated. Using a semiclassical
input-output theory for the interaction between an ensemble of four-level atoms and three optical
cavity fields coupled to the same spatial cavity mode, we derive the steady-state transmission
spectra of the probe field and discuss the dynamics of the intracavity field buildup. The
analytical and numerical results are in good agreement with recent experiments with ion Coulomb
crystals [M. Albert \textit{et al.}, Nature Photon. {\bf 5}, 633 (2011)].
\end{abstract}


\date{\today}

\maketitle

\section{Introduction}

Electromagnetically Induced Transparency (EIT) is a quantum interference phenomenon occurring when
two electromagnetic fields resonantly excite two different transitions sharing a common
state~\cite{harris97,lukin01,fleischhauer05}. An intense control field addressing one of the
transitions can substantially modify the linear dispersion and absorption of an atomic medium for a
weak probe field resonant with the second transition. Since its first observation by Boller
\textit{et al.}~\cite{boller91}, EIT has been successfully exploited for instance to control the
propagation of light pulses through an otherwise opaque medium for light storage and
retrieval~\cite{hau99,kash99,budker99,liu01,phillips01,turukin01,bigelow03,bajcsy03} and quantum
memories~\cite{chaneliere05,eisaman05,simon07,appel08,honda08,cviklinski08,choi08,zhaor09,zhaob09,lvovsky09}.
Besides providing a means for controlling the linear susceptibility of an atomic medium EIT
can also be exploited for generating strong optical
nonlinearities~\cite{schmidt96,harris99,lukin01,fleischhauer05}. For instance, in the four-level
atomic configuration such as the one depicted in Fig.~\ref{fig:scheme}b, the nonlinear
susceptibility of the medium can be strongly enhanced at the same time as the linear susceptibility
is suppressed. The large cross-Kerr effect between the probe field and a third
\textit{switching} field can then be used e.g. for high-efficiency photon
counting~\cite{imamoglu02}, all-optical switching~\cite{kang03,braje04} and nonlinear optics
at low-light levels~\cite{harris98,harris99,lukin01,fleischhauer05,ottaviani03}, nonclassical state
generation~\cite{lukin00nl} or the realization of strongly interacting photon gases~\cite{chang08,angelakis11}.

When such a nonlinear EIT medium is positioned in an optical cavity one first-of-all benefits from the
enhanced interaction of the ensemble with well-defined spatio-temporal field modes. This is of great value for enhancing the effective optical depth and for realizing high-efficiency quantum
memories~\cite{lukin00,dantan04,dantan06,gorshkov07} or Fock state quantum
filters~\cite{nikoghosyan10}. The EIT-induced reduction of the cavity linewidth~\cite{lukin98} can
also be used to increase the sensitivity of atomic magnetometers~\cite{budker99,scully92}, enhance cavity
optomechanical cooling processes~\cite{tan11,genes11,morigi11}, achieve lasing without
inversion~\cite{wu08lwi}, optical switching~\cite{nielsen11,vuletic11} or quantum state swapping~\cite{dantan11}.

On the other hand, the cavity EIT-interaction can be furthermore exploited to enhance the cross-Kerr
nonlinearity to investigate e.g. photon-blockade mechanisms~\cite{imamoglu97,grangier98,gheri99},
photon-photon interactions~\cite{werner99}, highly-entangled state generation~\cite{dantan06cpt} or
novel quantum phase transitions for light~\cite{hartmann06}.

Cavity EIT has been observed with atomic beams~\cite{muller97}, in cold and room temperature atomic
ensembles~\cite{hernandez07,wu08,laupetre11}, and even with single or few atoms in high-finesse optical
cavities~\cite{mucke10,kampschulte10}. Cavity EIT as well as EIT-based optical switching have also recently been observed with cold ion Coulomb crystals~\cite{albert11}.

We theoretically investigate here the
interaction between an ensemble of atoms with the four-level structure depicted in
Fig.~\ref{fig:scheme}b and three optical fields coupled to the same spatial cavity mode. Using a
semiclassical input-output theory we derive the atomic linear and nonlinear susceptibilities for
the probe field and its steady state transmission spectra as it experiences EIT or an EIT-based
cross-Kerr effect. We study in particular the effect of the transverse mode profiles of the fields
on the normal mode spectrum of the atom-cavity system as in the experiments of Ref.~\cite{albert11} and
compare with the more usual situation where only the probe field is coupled to the
cavity~\cite{muller97,hernandez07,wu08,mucke10,kampschulte10,laupetre11}.

The paper is organized as follows: in Sec.~\ref{sec:system} we introduce the system under consideration and
derive the equations of motion. We obtain the probe field susceptibility and transmission spectrum,
first when the probe and control fields interact with the atoms in an EIT situation in
Sec.~\ref{sec:EIT}, then when all three cavity fields interact simultaneously with the atomic
medium in Sec.~\ref{sec:switching}. Finally, all-optical switching at low-light levels using ion
Coulomb crystals as in the experiments of Ref.~\cite{albert11} is discussed in
Sec.~\ref{sec:discussion}.

\section{System considered and equations of motion}\label{sec:system}

\begin{figure}[h]
\includegraphics[width=0.4\columnwidth]{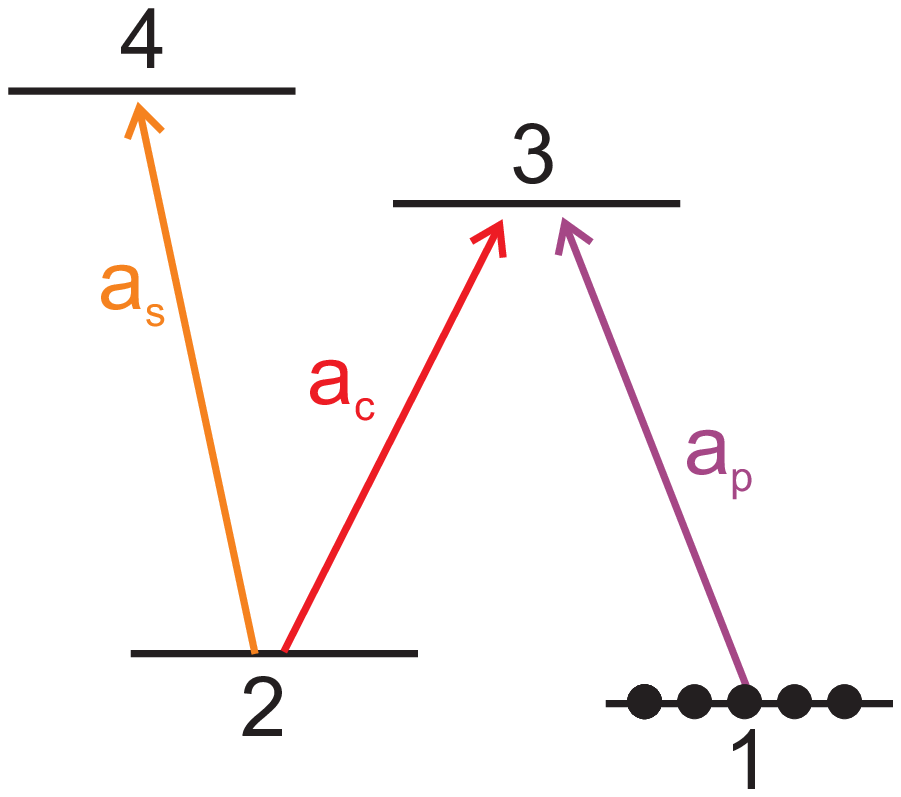}\hspace{0.2cm}\includegraphics[width=0.55\columnwidth]{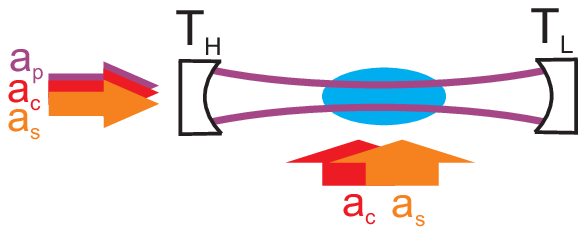}
\caption{(a) Atomic four-level structure considered. (b) System considered: an ensemble of four-level atoms, positioned in a linear optical cavity, interact with three optical fields. In the \textit{all-cavity} geometry all three fields are injected into the cavity and coupled to the same cavity mode. In an alternative \textit{standard} geometry, only the probe field $a_p$ is injected into the cavity, while the control and switching fields, $a_c$ and $a_s$, are free-propagating with waists much larger than that of the ensemble (see text for details).}
\label{fig:scheme}
\end{figure}

We consider an ensemble of $N_{tot}$ four-level atoms with the level structure depicted in
Fig.~\ref{fig:scheme}, where levels $|1\rangle$ and $|2\rangle$ are long-lived ground or metastable
states, and levels $|3\rangle$ and $|4\rangle$ are excited states. The atoms are enclosed in a
linear optical cavity where they interact with three optical fields: a \textit{probe} field on the
$|1\rangle\longrightarrow|3\rangle$ transition, a \textit{control} field on the
$|2\rangle\longrightarrow|3\rangle$ transition and a \textit{switching} field on the
$|2\rangle\longrightarrow|4\rangle$ transition. The cavity is assumed to have asymmetric mirror
transmissions, $T_H$ and $T_L$, with $T_H\gg T_L$, and, unless stated otherwise, all fields are
injected into the cavity through the high-transmission mirror. The annihilation operators of the
field modes in the Heisenberg picture and in the rotating frame are denoted by $\hat{a}_p$,
$\hat{a}_c$ and $\hat{a}_s$, respectively, with the convention
$[\hat{a}_{\alpha},\hat{a}_{\alpha}^{\dagger}]=1$, ($\alpha=p,c,s$).
$\hat{\sigma}_{\mu\nu}^{(j)}=|\mu\rangle\langle\nu|$ is the atomic operator associated with the $j-$th
ion positioned at $\textbf{r}_j$ ($\mu,\nu=1-4$). We assume that all three fields are resonant or
close to a resonance with the same spatial cavity mode, which we take to be the
fundamental Gaussian TEM$_{00}$ mode for simplicity. The interaction Hamiltonian in the rotating-wave
approximation and in the rotating frame is given by
\begin{align}
\label{eq:hamiltonian1}H_{af} =&-\hbar\sum_j g_p\Psi_p(\textbf{r}_j)\cos\left[k_{p}z_j+\varphi_p(\textbf{r}_j)\right]
\hat{a}_p\hat{\sigma}_{31}^{(j)}\\
\nonumber &-\hbar\sum_j g_c\Psi_c(\textbf{r}_j)\cos\left[k_{c}z_j+\varphi_c(\textbf{r}_j)\right]\hat{a}_c\hat{\sigma}_{32}^{(j)}\\
\nonumber &-\hbar\sum_jg_s\Psi_s(\textbf{r}_j)\cos\left[k_{s}z_j+\varphi_s(\textbf{r}_j)\right]\hat{a}_s\hat{\sigma}_{42}^{(j)}+\textrm{h.c.},
\end{align}
where the $g$'s are the maximal single-atom coupling strengths for the transitions considered, the
$\Psi$'s and $\varphi$'s are the fields' transverse mode functions and longitudinal mode phases,
respectively~\cite{kogelnik66}. Since our main goal is to discuss the effects related to the
spatial transverse structure of the fields we will consider for simplicity the situation where the length, $2L$, of the ensemble is much smaller than the Rayleigh range of the
cavity, as e.g. in the experiments of Ref.~\cite{albert11}. We thus neglect the longitudinal variations of the phases and the waists of the light fields over the length of
the ensemble. We therefore set the longitudinal mode phases to 0 and assume a Gaussian transverse structure of
the fields given by $\Psi_{\alpha}(r)=\exp(-r^2/w_{\alpha}^2)$ ($\alpha=p,c,s$), where $w_{\alpha}$ is the
cavity waist considered. We will in the following examine two situations:

\textit{(i)} The novel case in which all three fields have the same transverse
mode profile and the ensemble has a large radial extension as compared to the waists, like e.g. in
the experiments of~\cite{albert11}. We shall refer to this situation as the \textit{all-cavity}
case.

\textit{(ii)} The more usual situation in which the control and switching fields have a large
transverse intensity profile as compared to the extension of the ensemble. This would typically be the
case if these fields were interacting with the atoms not through the
cavity~\cite{hernandez07,wu08,mucke10,kampschulte10,laupetre11}, or for an ensemble radially confined to a region with dimension much smaller than the waists, as could be
obtained with e.g. with a string of atoms or two-component ion Coulomb crystals~\cite{hornekaer01}. We shall refer to such a situation, in which the transverse mode profiles of the fields can be ignored, as the \textit{standard} case.

We will in addition assume that, because of their motion, the atoms are ''warm" enough such that
they probe any field variation along the longitudinal standing-wave structures of the fields during
the characteristic time scales of the dynamics of the fields due their interactions with the atoms.
As discussed e.g. in~\cite{zimmer06,hansen07,lin09,wu10} and in Appendix~\ref{sec:appendix}, one can under these conditions assume
averaged longitudinal couplings $\bar{g}_{\alpha}= g_{\alpha}/\sqrt{2}$ $(\alpha=p,c,s)$. Keeping only the
transverse spatial dependence of the cavity modes, the Hamiltonian becomes
\begin{align} \nonumber H_{af}=&-\hbar\sum_j \bar{g}_p\Psi_p(r_j)
\hat{a}_p\hat{\sigma}_{31}^{(j)}+\bar{g}_c\Psi_c(r_j)\hat{a}_c\hat{\sigma}_{32}^{(j)}\\
&+\bar{g}_s\Psi_s(r_j)\hat{a}_s\hat{\sigma}_{42}^{(j)}+\textrm{h.c.}\label{eq:hamiltonian2}
\end{align}
For comparison the ''cold" atom situation where the atoms are well-localized with respect to the field
standing wave structures during the interaction is treated in Appendix~\ref{sec:appendix}.\\

The atom-field dynamics of the observable mean values can be standardly derived via
$\dot{o}=(1/i\hbar)\langle [\hat{o},H]\rangle$, where $o\equiv\langle\hat{o}\rangle$ is the mean
value of observable $\hat{o}$ and the total Hamiltonian $H=H_{a}+H_{f}+H_{af}$ is the sum of the
interaction Hamiltonian (\ref{eq:hamiltonian2}) and of the atomic and field Hamiltonians
\begin{eqnarray}
\nonumber H_a&=&-\hbar\sum_j\Delta_p\hat{\sigma}_{33}^{(j)}+(\Delta_p-\Delta_c)\hat{\sigma}_{22}^{(j)}+(\Delta_p-\Delta_c+\Delta_s)\hat{\sigma}_{44}^{(j)},\\
H_f&=&-\hbar\Delta^c_p\hat{a}_p^{\dagger}\hat{a}_p-\hbar\Delta^c_c\hat{a}_c^{\dagger}\hat{a}_c-\hbar\Delta^c_s\hat{a}_s^{\dagger}\hat{a}_s,
\end{eqnarray}
where $\Delta_{p}=\omega_{p}-\omega_{31}$, $\Delta_{c}=\omega_{c}-\omega_{32}$ and $\Delta_{s}=\omega_{s}-\omega_{42}$ are the one-photon detunings, and
$\Delta^c_{\alpha}=\omega_{\alpha}-\omega_{cav}$ are the cavity detunings between the
fields with frequency $\omega_{\alpha}$ and the cavity resonance frequency $\omega_{cav}$ $(\alpha=p,c,s)$. Denoting by $\gamma_{31}$, $\gamma_{32}$ and
$\gamma_{42}$ the spontaneous decay rates and introducing a phenomenological decay rate $\gamma_0$
for the ground-state coherence operators $\hat{\sigma}_{12}^{(j)}$ ($\gamma_0\ll\gamma_{31},\gamma_{32},\gamma_{42}$), one obtains the following set of coupled
differential equations
\begin{widetext}
\begin{eqnarray}
\label{eq:fullfirst}\dot{\sigma}_{12}^{(j)}&=&-(\gamma_0-i(\Delta_p-\Delta_c))\sigma_{12}^{(j)}-i\bar{g}_p\Psi_p(r_j)a_p\sigma_{32}^{(j)}+i\bar{g}_c\Psi_c(r_j)a_c^*\sigma_{13}^{(j)}+i\bar{g}_s\Psi_s(r_j)a_s^*\sigma_{14}^{(j)}\\
\dot{\sigma}_{13}^{(j)}&=&-(\gamma-i\Delta_p)\sigma_{13}^{(j)}+i\bar{g}_p\Psi_p(r_j)a_p(\sigma_{11}^{(j)}-\sigma_{33}^{(j)})+i\bar{g}_c\Psi_c(r_j)a_c\sigma_{12}^{(j)}\\
\dot{\sigma}_{14}^{(j)}&=&-(\gamma_s-i(\Delta_p-\Delta_c+\Delta_s))\sigma_{14}^{(j)}-i\bar{g}_p\Psi_p(r_j)a_p\sigma_{34}^{(j)}+i\bar{g}_s\Psi_s(r_j)a_s\sigma_{12}^{(j)}\\
\dot{\sigma}_{23}^{(j)}&=&-(\gamma-i\Delta_c)\sigma_{23}^{(j)}+i\bar{g}_c\Psi_c(r_j)a_c(\sigma_{22}^{(j)}-\sigma_{33}^{(j)})+i\bar{g}_p\Psi_p(r_j)a_p\sigma_{21}^{(j)}-i\bar{g}_s\psi_s(r_j)a_s\sigma_{43}^{(j)}\\
\dot{\sigma}_{24}^{(j)}&=&-(\gamma_s-i\Delta_s)\sigma_{24}^{(j)}-i\bar{g}_c\Psi_c(r_j)a_c\sigma_{34}^{(j)}+i\bar{g}_s\Psi_s(r_j)a_s(\sigma_{22}^{(j)}-\sigma_{44}^{(j)})\\
\dot{\sigma}_{34}^{(j)}&=&-(\gamma+\gamma_s-2\gamma_0-i(\Delta_s-\Delta_c))\sigma_{34}^{(j)}-i\bar{g}_c\Psi_c(r_j)a_c^*\sigma_{24}^{(j)}+i\bar{g}_s\Psi_s(r_j)a_s\sigma_{32}^{(j)}-i\bar{g}_p\Psi_p(r_j)a_p^*\sigma_{14}^{(j)}\\
\dot{\sigma}_{11}^{(j)}&=&\gamma_{31}\sigma_{33}^{(j)}-i\bar{g}_p\Psi_p(r_j)a_p\sigma_{31}^{(j)}+i\bar{g}_p\Psi_p(r_j)a_p^*\sigma_{13}^{(j)}\\
\dot{\sigma}_{22}^{(j)}&=&\gamma_{32}\sigma_{33}^{(j)}+\gamma_{42}\sigma_{44}^{(j)}-i\bar{g}_c\Psi_c(r_j)a_c\sigma_{32}^{(j)}+i\bar{g}_c\Psi_c(r_j)a_c^*\sigma_{23}^{(j)}
-i\bar{g}_s\Psi_s(r_j)a_s\sigma_{42}^{(j)}+i\bar{g}_s\Psi_s(r_j)a_s^*\sigma_{24}^{(j)}\\
\dot{\sigma}_{33}^{(j)}&=&-(\gamma_{31}+\gamma_{32})\sigma_{33}^{(j)}+i\bar{g}_p\Psi_p(r_j)a_p\sigma_{31}^{(j)}-i\bar{g}_p\Psi_p(r_j)a_p^*\sigma_{13}^{(j)}+i\bar{g}_c\Psi_c(r_j)a_c\sigma_{32}^{(j)}-i\bar{g}_c\Psi_c(r_j)a_c^*\sigma_{23}^{(j)}\\
\dot{\sigma}_{44}^{(j)}&=&-\gamma_{42}\sigma_{44}^{(j)}+i\bar{g}_s\Psi_s(r_j)a_s\sigma_{42}^{(j)}-i\bar{g}_s\Psi_s(r_j)a_s^*\sigma_{24}^{(j)}\\
\dot{a}_p&=&-(\kappa-i\Delta_p^c)a_p+i\sum_j\bar{g}_p\Psi_p(r_j)\sigma_{13}^{(j)}+\sqrt{\frac{2\kappa_H}{\tau}}a_p^{in}\\
\dot{a}_c&=&-(\kappa-i\Delta_c^c)a_c+i\sum_j\bar{g}_c\Psi_c(r_j)\sigma_{23}^{(j)}+\sqrt{\frac{2\kappa_H}{\tau}}a_c^{in}\\
\label{eq:fulllast}\dot{a}_s&=&-(\kappa-i\Delta_s^c)a_s+i\sum_j\bar{g}_s\Psi_s(r_j)\sigma_{24}^{(j)}+\sqrt{\frac{2\kappa_H}{\tau}}a_s^{in}
\end{eqnarray}
\end{widetext}
where $\gamma=(\gamma_{31}+\gamma_{32})/2+\gamma_0$, $\gamma_s=\gamma_{42}/2+\gamma_0$ and $\tau$ is the
cavity round-trip time. The input fields are denoted by $a^{in}_{\alpha}$ ($\alpha=p,c,s$). The total
cavity field decay rate (assumed equal for all fields for simplicity) is denoted by
$\kappa=\kappa_H+\kappa_T+\kappa_A$, where $\kappa_{H,L}=T_{H,L}/2\tau$ are the decay rates
corresponding to the mirrors' transmission and $\kappa_A=A/2\tau$ is a decay rate corresponding to
round-trip absorption losses $A$. While these absorption losses are not essential to the understanding of the physical mechanisms studied here, we include them for completeness as they very often affect experiments with high-finesse cavities~\cite{mucke10,kampschulte10,albert11}.

The previous set of equations can be solved numerically for any initial internal atomic state, input field pulses,
ensemble geometry and atomic distribution. We focus in the following on the situation in which all
the atoms are in state $|1\rangle$ initially and the probe field is much weaker than the control
and switching fields, so that one can perform a first-order expansion in the probe field to get analytical expressions for various quantities,
such as the probe susceptibility, its cavity transmission and reflection, the EIT buildup time, etc.

\section{Cavity Electromagnetically Induced Transparency}\label{sec:EIT}

\subsection{EIT regime}\label{sec:EITregime}

We first investigate the EIT situation where the atoms interact with both the probe and
control fields, but no switching field is injected into the cavity. The input probe and control
fields are abruptly switched on at time $t=0$ and have thereafter constant intensities. We place ourselves
in the weak probe regime, when $g_p|a_p|\ll g_c|a_c|$ and the intracavity photon number is much
smaller than the number of interacting atoms. All the atoms are then essentially in $|1\rangle$, and the
only non-zero atomic components at first order are the probe optical dipole $\sigma_{13}^{(j)}$ and the
ground-state coherence $\sigma_{12}^{(j)}$~\cite{fleischhauer05}. We also assume that the control field
is tuned to resonance with the $|2\rangle\longrightarrow|3\rangle$ transition ($\Delta_c=0$) and the cavity is
resonant with the $|2\rangle\longrightarrow|3\rangle$ transition, i.e.
$\Delta_p=\Delta_p^c=\delta=\Delta$. As the control field probes no atom, its intracavity amplitude
reaches its steady state value in a time $\kappa^{-1}$. Since we are interested in getting simple
analytical expressions for the steady state of the system and its dynamics over the typically
slower EIT buildup timescales, we can consider that the control field intracavity Rabi frequency is
constant and equal to its steady state value $\bar{\Omega}_c=\bar{g}_ca_c$. Equivalently, the control field can be turned on slightly before the probe pulse is applied. The relevant equations of motion
governing the evolution of the intracavity probe field are then
\begin{align}
\label{eq:a_p_EIT}\dot{a}_p&=-(\kappa-i\Delta)a_p+i\bar{g}_p\sum_j\Psi_p(r_j)\sigma_{13}^{(j)}+\sqrt{2\kappa_H/\tau}a_p^{in},\\
\label{eq:sigma_13_EIT}\dot{\sigma}_{13}^{(j)}&=-(\gamma-i\Delta)\sigma_{13}^{(j)}+i\bar{g}_p\Psi_p(r_j)a_p+i\bar{\Omega}_c\Psi_c(r_j)\sigma_{12}^{(j)},\\
\label{eq:sigma_12_EIT}
\dot{\sigma}_{12}^{(j)}&=-(\gamma_0-i\Delta)\sigma_{12}^{(j)}+i\bar{\Omega}_c^*\Psi_c(r_j)\sigma_{13}^{(j)}.\end{align}

\subsection{Steady state susceptibility}
These equations can be readily solved in steady state to obtain the mean value of the intracavity probe field,
\begin{equation}\label{eq:a_psteady}
a_p=\frac{\sqrt{2\kappa_H/\tau}a_p^{in}}{\kappa-i\Delta-i\chi_{EIT}},
\end{equation}
where the EIT susceptibility is given by
\begin{eqnarray}
\label{eq:chi_EIT}\chi_{EIT}=\sum_j\frac{i\bar{g}_p^2\Psi_p^2(r_j)}{\gamma-i\Delta+\frac{|\bar{\Omega}|_c^2}{\gamma_0-i\Delta}\Psi_c^2(r_j)}.
\end{eqnarray}
For large ensembles of atoms with non-correlated positions, we can reformulate Eq.~(\ref{eq:chi_EIT}) in terms of the local atomic density $\rho(\textbf{r})$. Assuming the
same transverse profiles for the control and probe fields $\Psi_p(r)=\Psi_c(r)=\exp(-r^2/w^2)$,
Eq.~(\ref{eq:chi_EIT}) can then be recasted in
\begin{eqnarray} \chi_{EIT}&=&\frac{ig_p^2}{2}\int_V d\textbf{r}\;\rho(\textbf{r})
\frac{e^{-2r^2/w^2}}{\gamma-i\Delta+\frac{\Omega_c^2/2}{\gamma_0-i\delta}e^{-2r^2/w^2}}.\end{eqnarray} where $\Omega_c=g_c|a_c|$.
For an ensemble with a uniform atomic density, such as ion Coulomb crystals in a linear radiofrequency trap as used in~\cite{herskind09,dantan09,albert11,albert11pra}, and with a large radial extension as compared to the cavity
waist (see e.g. \cite{herskind09,dantan09,albert11,albert11pra}), the integral can be calculated analytically and yields a susceptibility
\begin{eqnarray}\chi_{EIT}&=&\frac{ig_p^2N}{\gamma-i\Delta}\frac{\ln(1+\Theta)}{\Theta},\label{eq:chi_EITall}\end{eqnarray}
where
\begin{equation}\label{eq:Theta}\Theta=\frac{\Omega_c^2/2}{(\gamma-i\Delta)(\gamma_0-i\Delta)}\end{equation}
is an effective saturation parameter for the two-photon transition and $N=\rho \frac{\pi w^2}{2}L$
is the effective number of atoms defined in~\cite{herskind09,albert11pra}. This result can be compared to the
\textit{standard} situation in which the control field waist is much larger than that of the probe
field~\cite{harris97,fleischhauer05,banacloche09}
\begin{equation}\chi_{EIT}^{st}=\frac{ig_p^2N}{\gamma-i\Delta}\frac{1}{1+\Theta}\label{eq:chi_EITst}\end{equation}

\begin{figure}[h]
\includegraphics[width=\columnwidth]{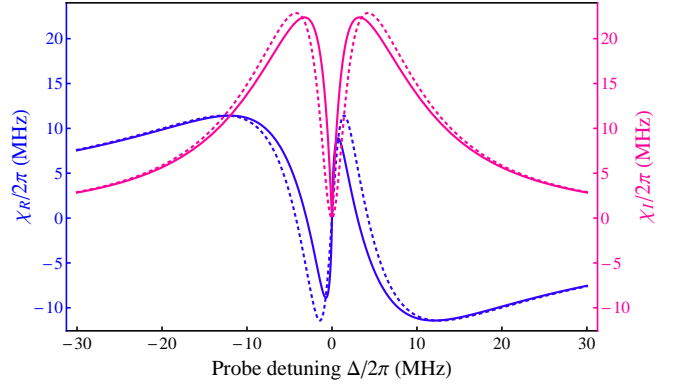}
\caption{(Color online) Real (blue) and imaginary (red) parts of the probe first-order
susceptibility as a function of the probe detuning $\Delta$ (solid line: \textit{all-cavity} EIT
[Eq.~(\ref{eq:chi_EITall})], dashed line: \textit{standard} EIT [Eq.~(\ref{eq:chi_EITst})]). The parameters,
$(g_p\sqrt{N},\gamma,\gamma_0,\Omega_c)=2\pi\times(16,11.2,6\times 10^{-4},6)$ MHz, are similar to those used in the experiments of Ref.~\cite{albert11}.}
\label{fig:eit_suscep}
\end{figure}

The real and imaginary part of these susceptibilities are plotted in Fig.~\ref{fig:eit_suscep} for
typical parameters used in the experiments with ion Coulomb crystals of~\cite{albert11}. They show
the typical transparency window in the absorption profile and the rapid change in dispersion around
two-photon resonance. In comparison with the standard one, the all-cavity susceptibility clearly shows non-Lorentzian lineshapes, as expected from its different dependence with respect to $\Theta$ [Eqs.~(\ref{eq:chi_EITall}) and (\ref{eq:chi_EITst})].

From Eq.~(\ref{eq:a_psteady}) and the input-output relations
\begin{equation}a_p^{ref}=\sqrt{2\kappa_T\tau}a_p-a_p^{in},\hspace{0.2cm}a_p^{tr}=\sqrt{2\kappa_L\tau}a_p,\end{equation} the steady state cavity transmission and reflection for the probe field are given by
\begin{eqnarray} \emph{T}&\equiv &\left|\frac{a_p^{tr}}{a_p^{in}}\right|^2=\left|\frac{2\sqrt{\kappa_H\kappa_L}}{\kappa_H+\kappa_L+\kappa_A-i\Delta-i\chi}\right|^2,\\
\emph{R}&\equiv &\left|\frac{a_p^{ref}}{a_p^{in}}\right|^2=\left|\frac{\kappa_H-\kappa_L-\kappa_A+i\Delta+i\chi}{\kappa_H+\kappa_L+\kappa_A-i\Delta-i\chi}\right|^2.\end{eqnarray}
Using Eqs.~(\ref{eq:chi_EITall},\ref{eq:chi_EITst}) one can then compute the normal mode spectrum
of the probe field transmission. In the collective strong coupling regime, when
$g_p\sqrt{N}>\kappa,\gamma$, one
expects three normal modes in the transmission spectrum: two modes at probe detunings
$\pm\sqrt{g_p^2N+\Omega_c^2/2}$, corresponding to the two-level ensemble modes $\pm g_p\sqrt{N}$
shifted by the presence of the control field, and one mode at zero-detuning for the probe (two-photon
resonance here), corresponding to the cavity EIT resonance~\cite{wu08,banacloche09}.
\begin{figure}[h]
\includegraphics[width=\columnwidth]{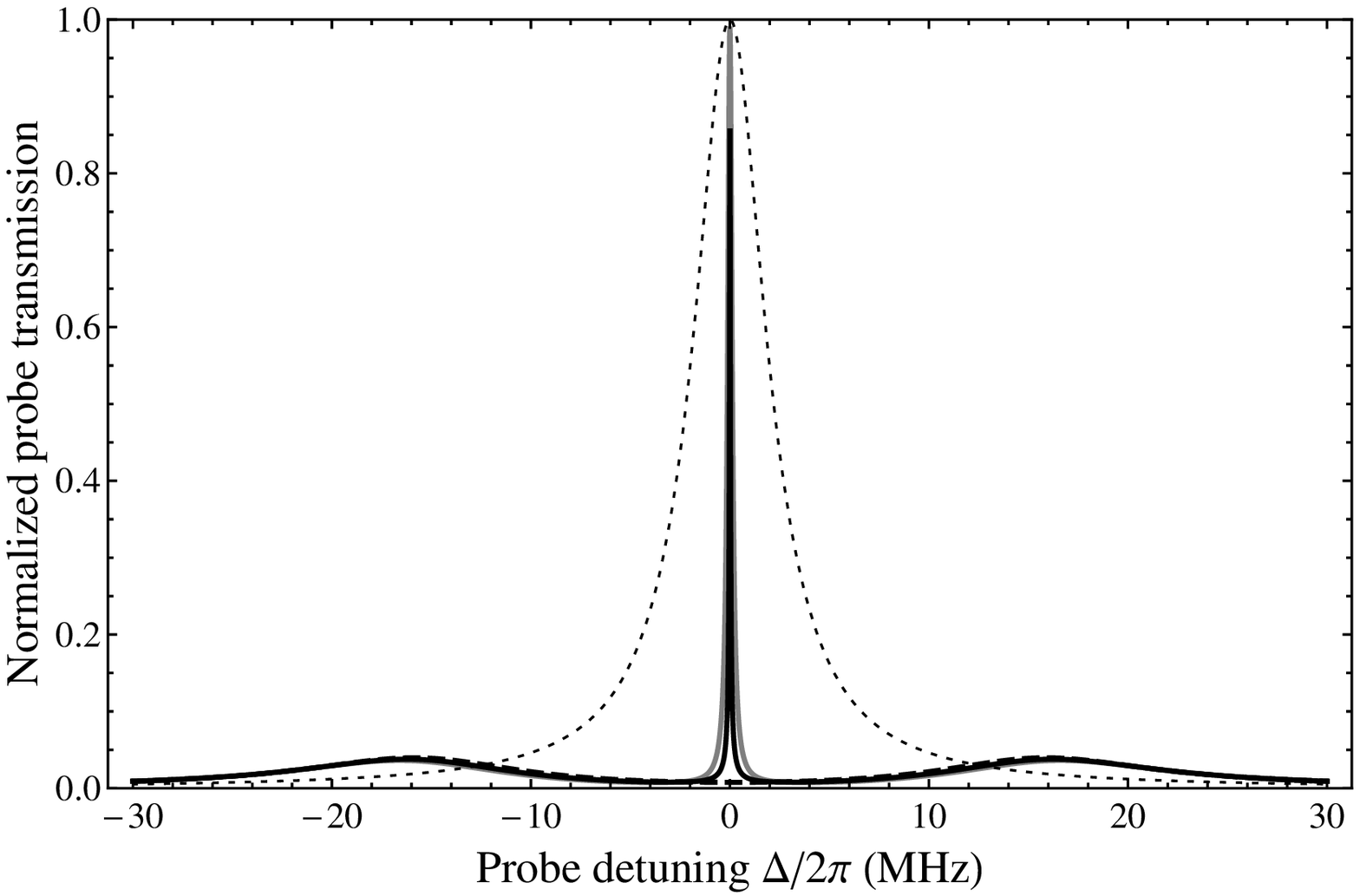}
\includegraphics[width=\columnwidth]{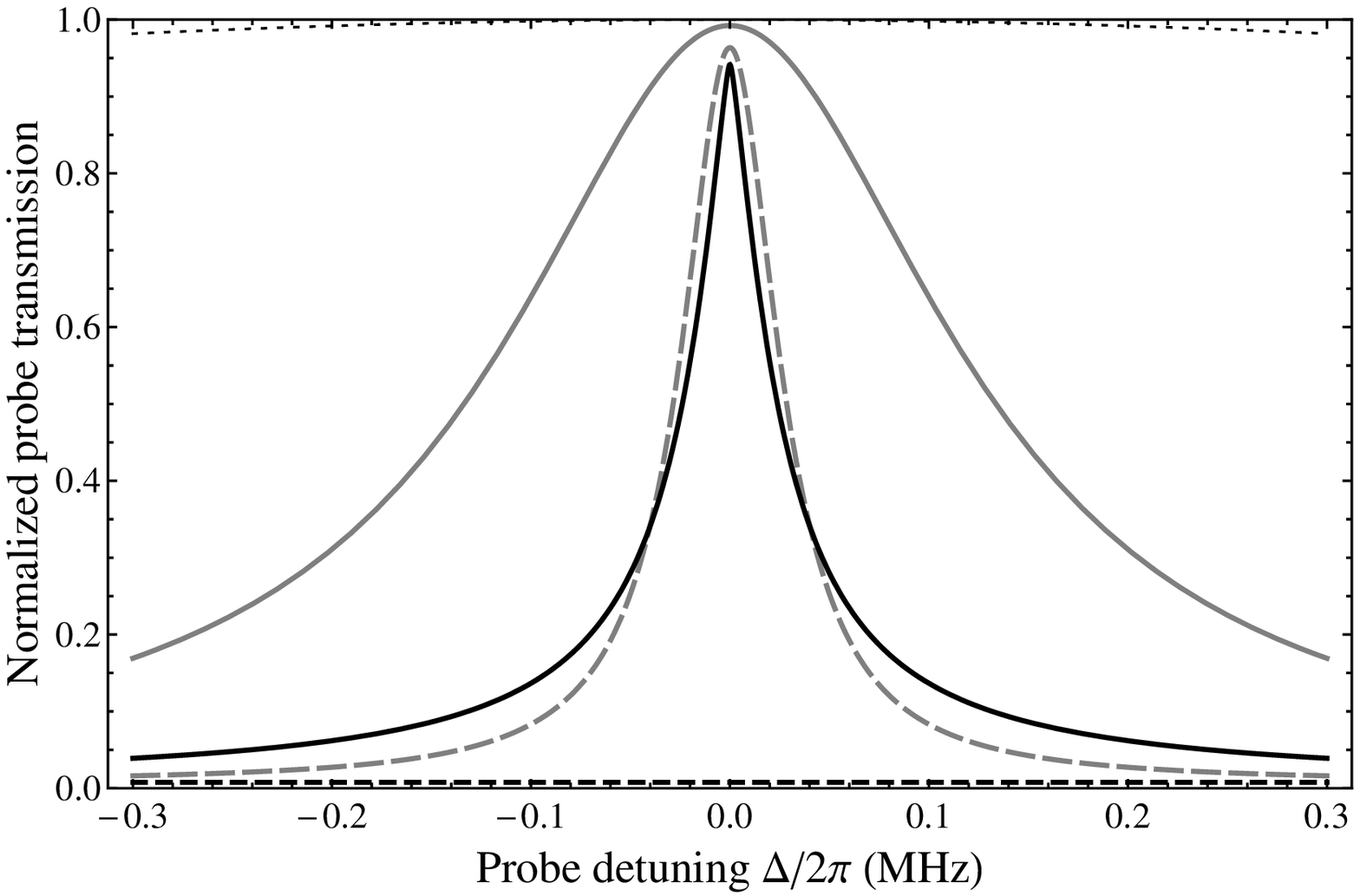}
\caption{Upper panel: Probe field normalized transmission spectra for an empty cavity (dotted line), a cavity
containing a uniform density ensemble interacting with the probe field only (dashed grey line) and a cavity
containing a uniform density ensemble interacting with a probe and a control field in an
\textit{all-cavity} (solid black line) and \textit{standard} EIT situation (gray solid/dashed lines). Parameters: $(g_p\sqrt{N},\gamma,\gamma_0,\Omega_c,\kappa)=2\pi\times(16,11.2,6\times 10^{-4},6,2.2)$ MHz. The solid and dashed grey curves show the \textit{standard} EIT situation with control field Rabi frequencies $\Omega_c=(2\pi)6$ MHz and $\Omega_c=\Omega_{c,\textrm{eff}}=(2\pi)6/2.2$ MHz, respectively. Lower panel: Spectra enlarged
around $\Delta=0$. As discussed in the text, the latter value of the effective Rabi frequency was chosen to illustrate the lineshape difference in a situation when the \textit{all-cavity} and \textit{standard} EIT resonance curves have comparable halfwidths at half-maximum.} \label{fig:eit_rabi_trans}
\end{figure}

This is illustrated in Fig.~\ref{fig:eit_rabi_trans}, where the probe transmission spectra, normalized to the bare cavity resonant value \begin{equation}\label{eq:T_0} \emph{T}_0=\left|\frac{\kappa}{\kappa-i\Delta-i\chi}\right|^2,\end{equation} is represented for the cases of \textit{(i)} an empty cavity ($\chi=0$), \textit{(ii)} a cavity
containing a uniform density ensemble interacting with the probe field only
($\chi=ig_p^2N/(\gamma-i\Delta)$) and a cavity containing a uniform density ensemble interacting with a
probe and a control field in an \textit{all-cavity} \textit{(iii)} and \textit{standard} \textit{(iv)} EIT situation, for which the probe susceptibility is given by (\ref{eq:chi_EITall}) and (\ref{eq:chi_EITst}), respectively. One observes indeed that the value of the cavity transmission in presence of EIT is restored to close to the bare cavity resonant value in a narrow frequency window around resonance. The width of the central EIT feature can be calculated by expanding the transmission around
two-photon resonance. In the \textit{standard} case and in the regime considered previously ($\gamma\gamma_0\ll\Omega_c^2<g_p^2N$), one
finds that the probe transmission is Lorentzian-shaped around $\Delta=0$ (see e.g.~\cite{lukin98,fleischhauer05})
\begin{equation}
\emph{T}\varpropto \left|\frac{1}{\kappa+\gamma_0\frac{g_p^2N}{\Omega_c^2/2}-i\Delta\frac{g_p^2N}{\Omega_c^2/2}}\right|^2\varpropto\frac{1}{|\kappa_{EIT}-i\Delta|^2}
\end{equation}
The interaction thus emulates a cavity with an effective halfwidth
\begin{equation}\kappa_{EIT}=\gamma_0+\kappa\frac{\Omega_c^2/2}{g_p^2N}\label{eq:kappa_EIT}\end{equation} which is
smaller than the bare cavity halfwidth $\kappa$ when $g_p\sqrt{N}>\Omega_c$. The analysis is a bit
more complicated in the \textit{all-cavity} case, due to the complex dependence of the
susceptibility with the saturation parameter $\Theta$ and the non-Lorentzian profile (as can be
seen e.g. from Fig.~\ref{fig:eit_rabi_trans}b).

The different dependence of the susceptibility with the effective saturation parameter $\Theta$ makes it in general impossible to define an effective control field Rabi frequency in the \textit{all-cavity} situation which would give the same susceptibility or transmission as in the \textit{standard} case. However, if one is interested in comparing situations in which the EIT resonance features have similar widths, one can perform a similar expansion of the normalized transmission given by Eqs.~(\ref{eq:chi_EITall}) and (\ref{eq:T_0}) around $\Delta=0$ and define an EIT resonance width also in the \textit{all-cavity} situation. In the regime $\Omega_c^2\gg\gamma\gamma_0$, the \textit{all-cavity} EIT resonance width matches the \textit{standard} EIT one with an effective control field Rabi frequency $\Omega_{c,\textrm{eff}}\simeq \Omega_c/[(\sqrt{2\pi^2+4(\ln 2C)^2}-\pi)/(\pi^2/2+2(\ln 2C)^2)]^{1/2}$, where $C=g_N^2/2\kappa\gamma$. For the parameters of Fig.~\ref{fig:eit_rabi_trans}, the scaling factor for the effective Rabi frequency is $\sim 2.2$ for instance. On the other hand, if one was interested in comparing the minimum absorption level on two-photon resonance , one could define an effective control Rabi frequency as $\Omega_{c,\textrm{eff}}'\sim\Omega_c/\sqrt{\ln(\Omega_c^2/\gamma\gamma_0)}$, in the EIT regime where $\Omega_c^2\gg\gamma\gamma_0$ and for a given value of $\Omega_c$. The resonant absorption in the \textit{all-cavity} situation with a control field Rabi frequency $\Omega_c$ could then be effectively compared to that of a \textit{standard} situation in which the maximal Rabi frequency has been scaled by a factor $\sim\sqrt{\ln(\Omega_c^2/\gamma\gamma_0)}$.

\begin{figure}[h]
\includegraphics[width=\columnwidth]{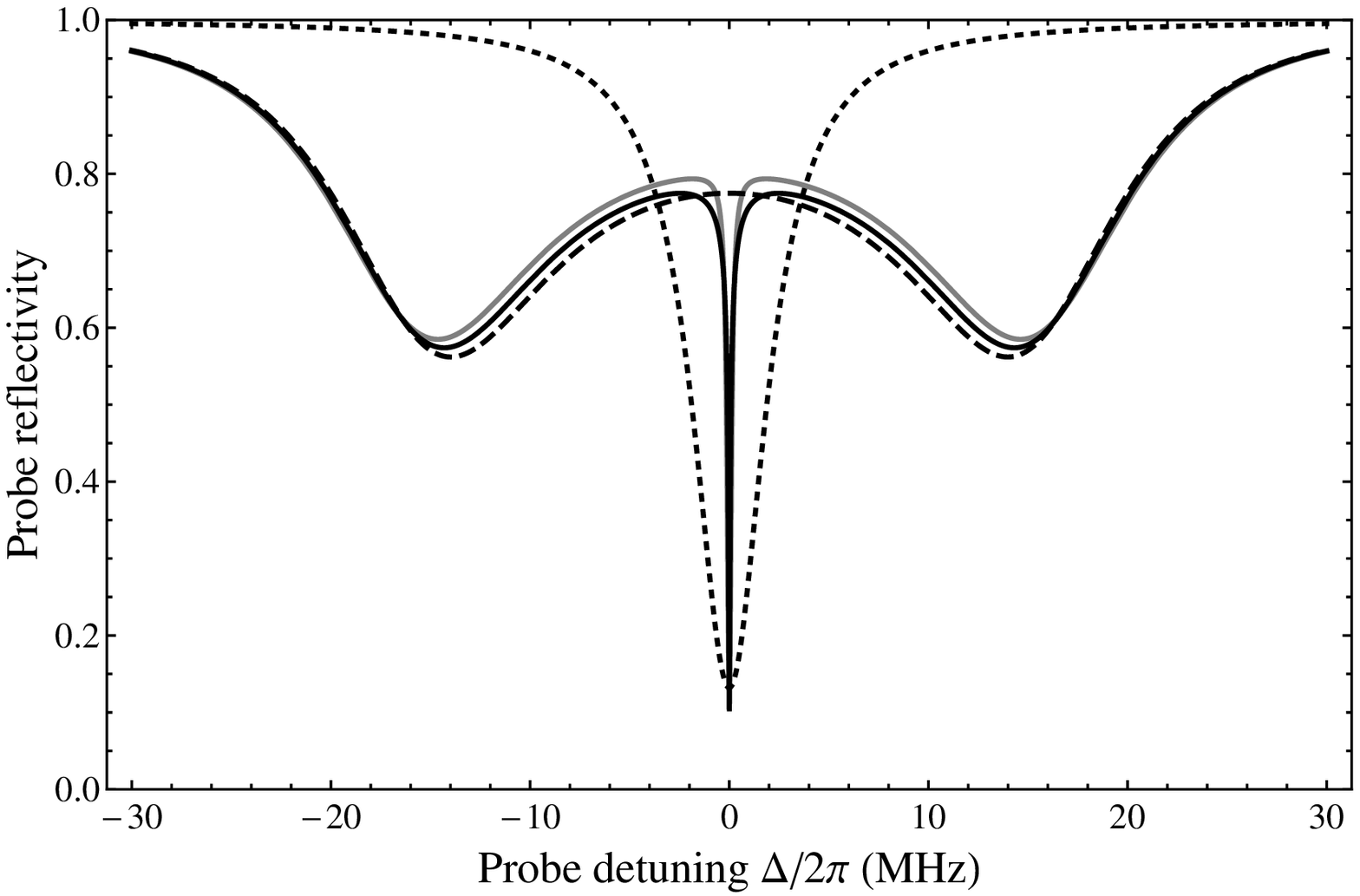}
\includegraphics[width=\columnwidth]{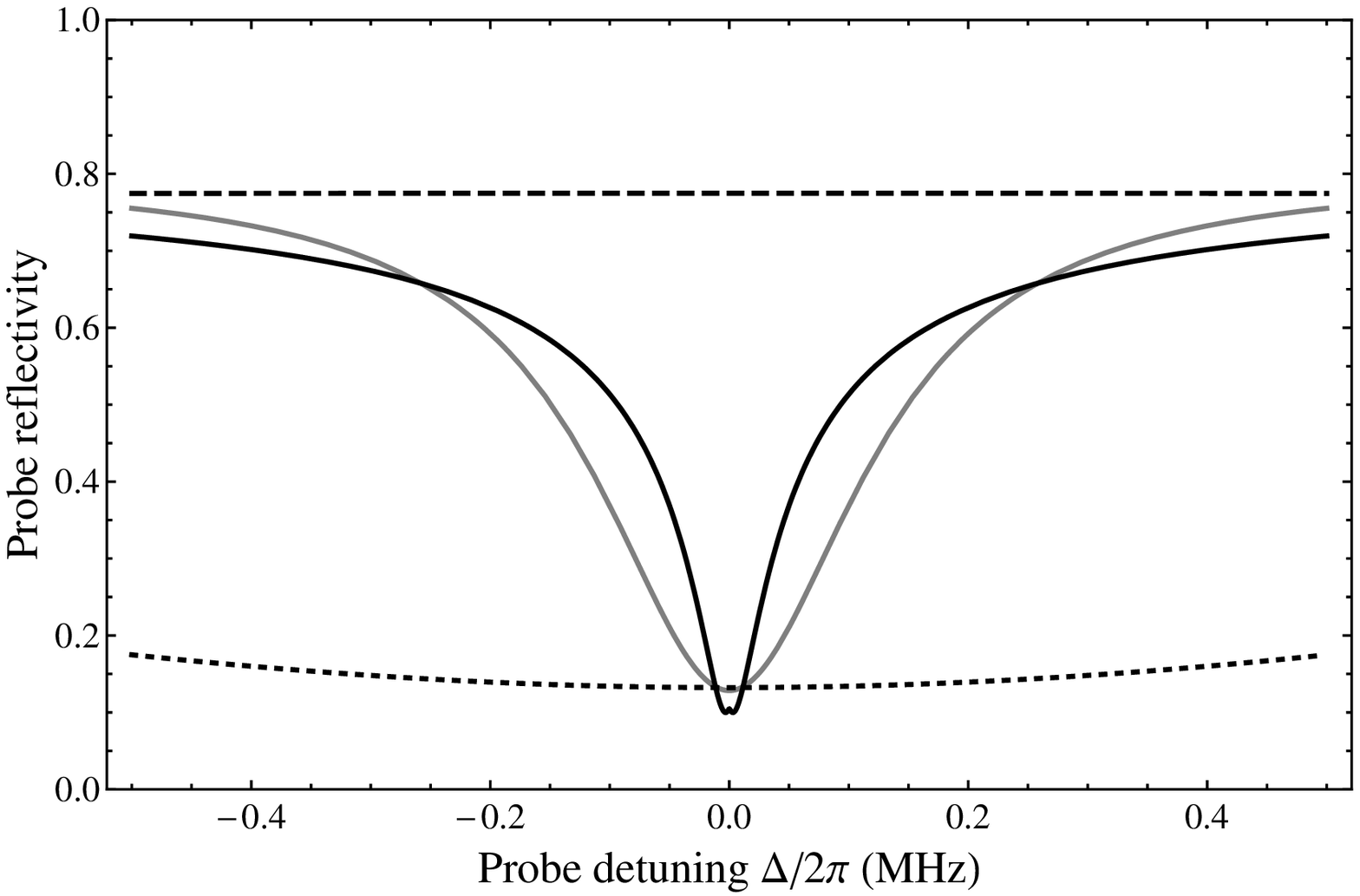}
\caption{Reflection spectra for the same configurations and parameters as in
Fig.~\ref{fig:eit_rabi_trans}, for a cavity with non negligible roundtrip absorption losses [$(\kappa_H,\kappa_L,\kappa_A)=(2\pi)\times(1.5,0,0.7)$ MHz].} \label{fig:eit_rabi_ref}
\end{figure}
Fig.~\ref{fig:eit_rabi_ref} shows the corresponding reflectivity spectra for a cavity having non negligible roundtrip absorption losses, as observed in the experiments of~\cite{albert11}. In general, since the reflected field results from the interference between the input field and the intracavity field, the reflectivity levels have a slightly more complex dependence on the atomic absorption and the cavity losses. The reflectivity spectrum exhibits nonetheless the same qualitative features as the transmission, with two normal modes at frequencies $\pm\sqrt{g_p^2N+\Omega_c^2/2}$ and a third one at zero-two-photon detuning corresponding to the reduction of atomic absorption due to the EIT effect. Effective control field Rabi frequencies can also be defined in a similar fashion as previously, as shown e.g. in~\cite{albertthesis,albert11}.

\subsection{Dynamics}

In this section we focus on the dynamics towards reaching the steady state during a resonant EIT interaction ($\Delta=0$).
Assuming again a constant control field Rabi frequency and performing a Laplace transform of
Eqs.~(\ref{eq:a_p_EIT}), (\ref{eq:sigma_13_EIT}) and (\ref{eq:sigma_12_EIT}) yields the following equations
\begin{eqnarray*}
(\kappa+s)a_p[s]&=&i\sum_j\bar{g}_p\Psi_p(r_j)\sigma_{13}^{(j)}[s]+\sqrt{\frac{2\kappa_H}{\tau}}a_p^{in}[s]\\
(\gamma+s)\sigma_{13}^{(j)}[s]&=&i\bar{g}_p\Psi_p(r_j)a_p[s]+i\bar{\Omega}_c\Psi_c(r_j)\sigma_{12}^{(j)}[s]\\
(\gamma_0+s)\sigma_{12}^{(j)}[s]&=&i\bar{\Omega}_c^*\Psi_c(r_j)\sigma_{13}^{(j)}[s]
\end{eqnarray*}
where the Laplace transform of $f$ is defined by
\begin{equation}
f[s]=\mathcal{L}[f(t)]=\int_0^{\infty}dt\;e^{-st}f(t)
\end{equation}
and we have assumed the initial conditions $a_p(0)=\sigma_{12}^{(j)}(0)=\sigma_{13}^{(j)}(0)=0$. These equations allow for extracting the Laplace transform of the intracavity probe field amplitude
\begin{equation}
a_p[s]=\frac{\sqrt{2\kappa_H/\tau}a_p^{in}[s]}{\kappa+s+\sum_j\frac{\bar{g}_p^2\Psi_p^2(r_j)}{\gamma+s+|\bar{\Omega_c}|^2/(\gamma_0+s)}}
\label{eq:a_plaplace}
\end{equation}
and calculate its time evolution by performing the inverse Laplace transform. It is however instructing to look at the
dynamics in the adiabatic limit in which the effective cavity linewidth
emulated by the EIT medium is smaller than the bare cavity linewidth and the dipole decay rate, i.e. $\kappa_{EIT}<\kappa,\gamma$. In this limit it can be shown that the intracavity field and the optical coherence
adiabatically both follow the ground state coherence, which evolves at a rate $\kappa_{EIT}$. In the \textit{standard} case, from Eq.~(\ref{eq:a_plaplace}), one finds that the intracavity field amplitude increases
exponentially with a time constant $1/\kappa_{EIT}$, consistently with the steady state spectrum
analysis of the previous section. In the \textit{all-cavity} case the inverse Laplace transform has to
be calculated numerically. It yields a non-exponential increase in the intracavity field intensity
occurring on a timescale approximately given by $1/\kappa_{EIT}$, with $\Omega_c$ scaled as in the
previous section.

\begin{figure}[h]
\includegraphics[width=\columnwidth]{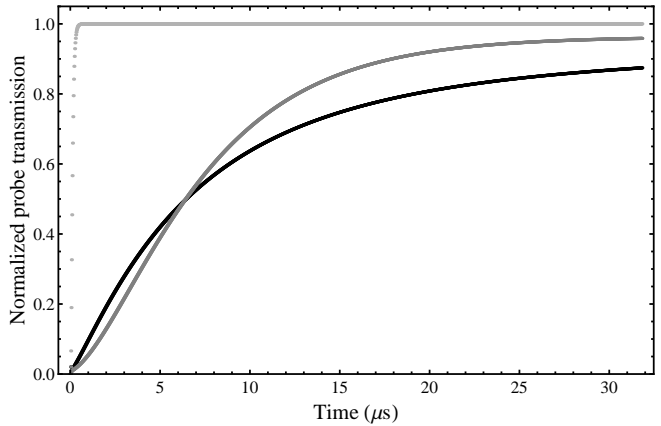}
\caption{Probe field normalized transmission as a function of time (solid black line: \textit{all-cavity} case,
solid grey line: \textit{standard} case, dotted grey line: empty cavity) for a resonant probe field ($\Delta=0$). Parameters as in
Fig.~\ref{fig:eit_rabi_trans}. The control field Rabi frequency is scaled by a factor 2.2 in the \textit{standard} situation.} \label{fig:eit_dynamics}
\end{figure}

Figure~\ref{fig:eit_dynamics} shows the time evolution of the normalized probe transmission in the two
situations discussed, for the same parameters as in the previous section and for an input probe
pulse abruptly switched on at $t=0$ and a constant control field. For comparison the (much faster) bare cavity response is also shown.

\section{Optical switching}\label{sec:switching}

We now turn to the all-optical switching situation, in which the transition
$|2\rangle\longleftrightarrow|4\rangle$ is addressed by the switching field
$\hat{a}_s$, while the control and probe fields are in an EIT situation. When the switching field is detuned from atomic resonance ($|\Delta_s|\gg\gamma_s$) and weak enough such that the absorption to level $|4\rangle$ is negligible, its main effect is to light-shift level $|2\rangle$, thereby changing the bare EIT resonance condition for the control and probe fields. When the light-shift becomes comparable or greater than the width of the cavity EIT window, the transmission of the probe field is inhibited, as the cavity is switched off resonance by the presence of the switching field.

\subsection{Probe susceptibility}

\begin{figure}[h]
\includegraphics[width=\columnwidth]{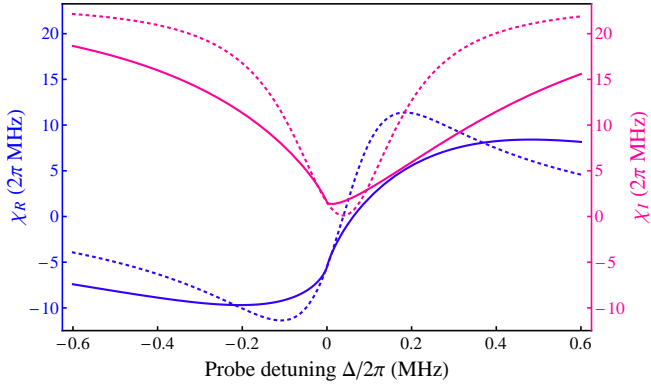}
\caption{(Color online) Optical switching susceptibility: real (blue) and imaginary (red) parts around two-photon
resonance. The solid line shows the \textit{all-cavity} switching for $(g_p\sqrt{N},\gamma,\kappa,\gamma_0,\Omega_c,\gamma_s,\Delta_s,\Omega_s)=2\pi\times
(16,11.2,2.2,0.0006,4,11,4300,40)$ MHz. The dashed line shows the \textit{standard} switching situation for the same parameters, but with the control and switching field Rabi frequencies scaled by a factor 2.2 for comparison.} \label{fig:switching_suscep}
\end{figure}

We assume again that almost all the atoms stay in $|1\rangle$ and that the control and switching field intracavity Rabi frequencies have reached their steady state values, $\bar{\Omega}_c=\bar{g}_ca_c$ and $\bar{\Omega}_s=\bar{g}_sa_s$, when the probe is injected. Performing a first-order treatment in the
probe field, the equations of motion for the non-zero coherences are given by
\begin{align}
\dot{\sigma}_{13}^{(j)}&=-(\gamma-i\Delta)\sigma_{13}^{(j)}+i\bar{g}_p\Psi_p(r_j)a_p+i\bar{\Omega}_c\Psi_c(r_j)\sigma_{12}^{(j)},\label{eq:sigma13switching}\\
\label{eq:sigma12switching}
\dot{\sigma}_{12}^{(j)}&=-(\gamma_0-i\Delta)\sigma_{12}^{(j)}+i\bar{\Omega}_c^*\Psi_c(r_j)
\sigma_{13}^{(j)}+i\bar{\Omega}_s^*\Psi_s(r_j)\sigma_{14}^{(j)},\\\label{eq:sigma14switching}
\dot{\sigma}_{14}^{(j)}&=-(\gamma_s-i\Delta_s-i\Delta)\sigma_{14}^{(j)}+i\bar{\Omega}_s\Psi_s(r_j)
\sigma_{12}^{(j)}.
\end{align}
Solving Eqs.~(\ref{eq:sigma13switching}), (\ref{eq:sigma12switching}) and (\ref{eq:sigma14switching}) in steady state readily yields a
mean intracavity probe field amplitude of the form (\ref{eq:a_psteady}), with a susceptibility
\begin{eqnarray}
\chi_{SW}&=&\sum_ji\bar{g}_p^2\Psi_p^2(r_j)\left[\gamma-i\Delta+\frac{\Omega_c^2\Psi_c^2(r_j)/2}{\gamma_0-i\Delta+\frac{\Omega_s^2\Psi_s^2(r_j)/2}{\gamma_s-i\Delta_s-i\Delta}}\right]^{-1}
\end{eqnarray}
with $\Omega_s=g_s|a_s|$. For a uniform density medium with an extension larger than the cavity waist and for fields with identical transverse profiles, one gets an analytical expression for the susceptibility
\begin{eqnarray}
\chi_{SW}&=&\frac{ig_p^2N}{\gamma-i\Delta}\left[\frac{\Theta\ln(1+\Theta+\Theta_s)}{(\Theta+\Theta_s)^2}+\frac{\Theta_s}{\Theta+\Theta_s}\right],
\end{eqnarray}
where \begin{equation}\Theta_s=\frac{\Omega_s^2/2}{(\gamma_s-i\Delta_s-i\Delta)(\gamma_0-i\Delta)}\end{equation} is defined analogously to the effective EIT saturation parameter for the probe. This
susceptibility can again be compared to that of the \textit{standard} case where the control and switching
fields have waists much larger than that of the probe
\begin{equation}
\chi_{SW}^{st}=\frac{ig_p^2N}{\gamma-i\Delta}\frac{1}{1+\Theta/(1+\Theta_s)}.
\end{equation}

\subsection{Probe field transmission spectrum}
\begin{figure}[h]
\includegraphics[width=\columnwidth]{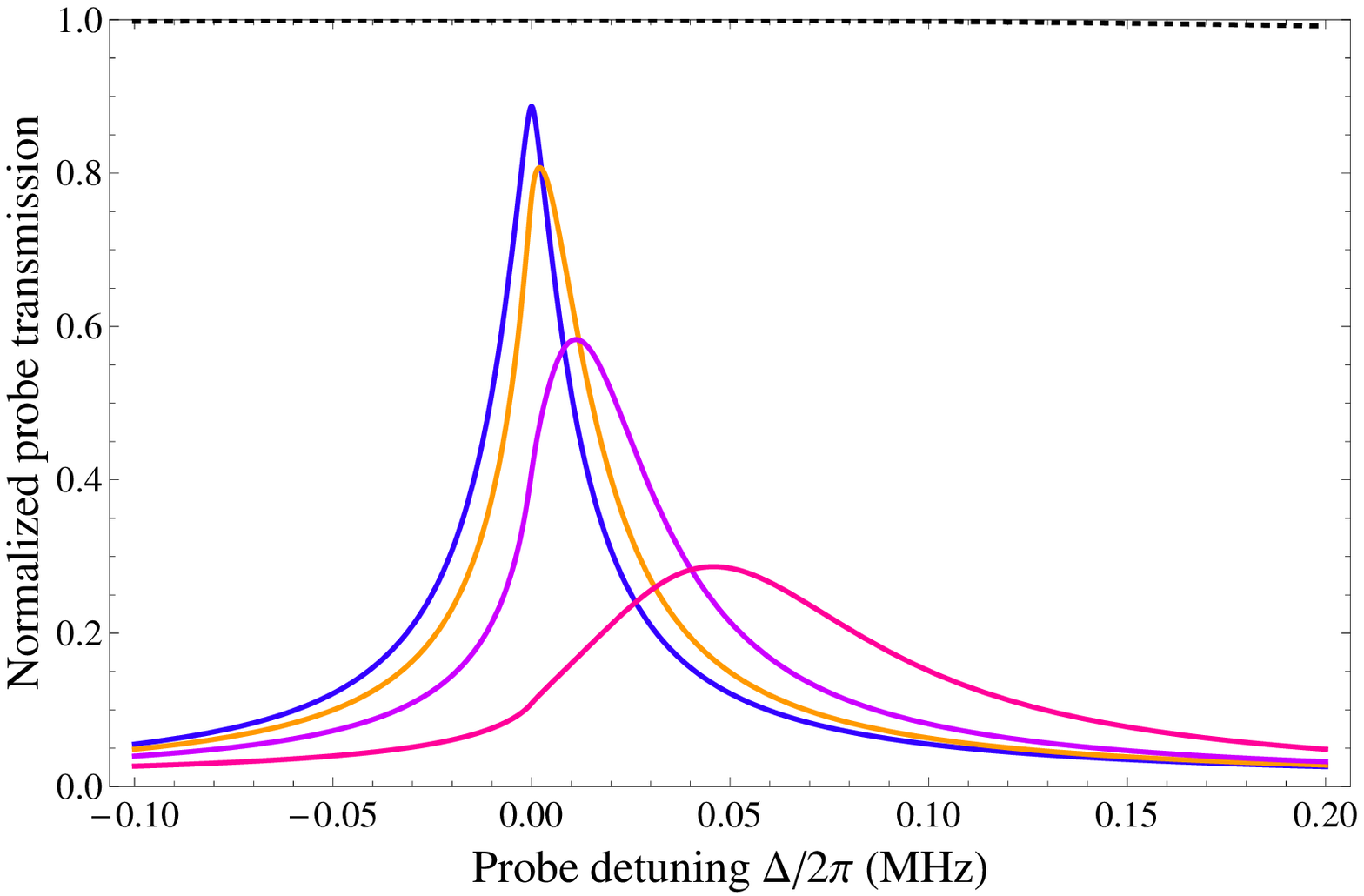}
\includegraphics[width=\columnwidth]{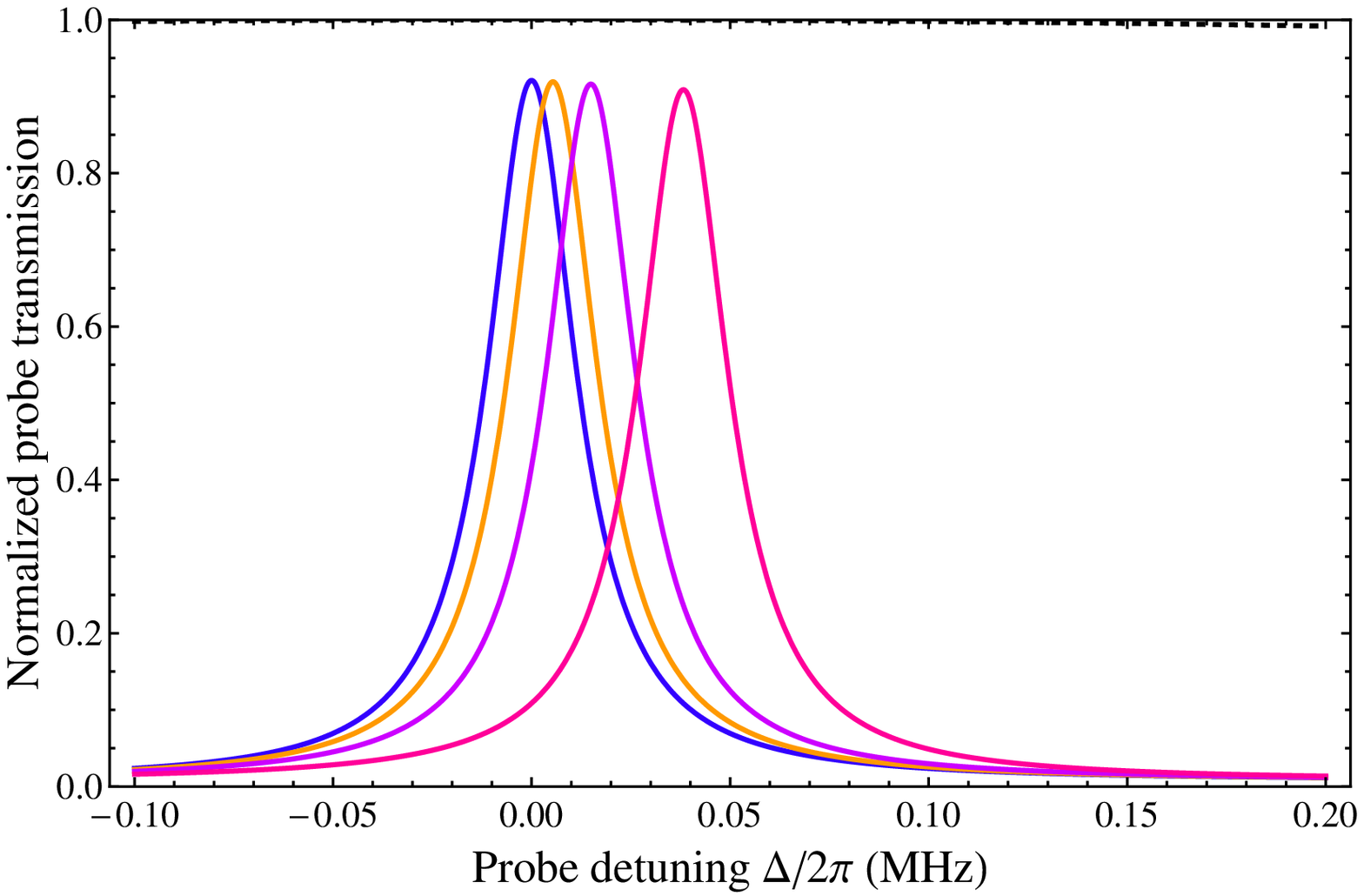}
\caption{(Color online) Probe field normalized transmission spectrum for different switching field Rabi frequencies $\Omega_s$: (a) \textit{All-cavity} switching. $\Omega_s=2\pi\times(0,15,25,40)$ MHz from left to right. Other parameters as in Fig.~\ref{fig:switching_suscep}. (b)
\textit{Standard} switching for the same parameters but with Rabi frequencies $\Omega_c$ and $\Omega_s$ scaled by a factor 2.2.}
\label{fig:switching_trans}
\end{figure}

The real and imaginary part of both susceptibilities are shown in Fig.~\ref{fig:switching_suscep} for typical parameter values taken from~\cite{albert11}. As expected, in the \textit{standard} case, the effect of the switching field is to shift the position of the EIT resonance by an amount that corresponds to the AC Stark shift of level $|2\rangle$. The effect is more complex in the \textit{all-cavity} case, as the shift for each atom depends on its radial position, thus leading to an asymmetric frequency behavior of the absorption and dispersion around two-photon resonance. This effects are manifest on the probe field transmission spectra, which are shown in Fig.~\ref{fig:switching_trans} for different switching field intensities. While, in the \textit{standard} configuration, the probe transmission profile is shifted away from the bare two-photon resonance without too much distortion as the switching field intensity increases, it is substantially distorted in the \textit{all-cavity} case due to the different AC Stark shifts experienced by atoms at different radial positions. The accuracy of these analytical expressions for the susceptibility and transmission have been checked by numerically solving Eqs.~(\ref{eq:fullfirst})-(\ref{eq:fulllast}). These findings are also in good agreement with the experimental observations of Ref.~\cite{albert11}.

\section{Low-light level optical switching with ion Coulomb
crystals in cavities}\label{sec:discussion}

We now turn to the prospects of achieving low-light optical switching of a single-photon probe field using ion Coulomb crystals in an optical cavity and base our practical discussion on the parameters of Refs.~\cite{herskind09,albert11}. We assume an asymmetric linear cavity geometry similar to that described in~\cite{albert11}, with length $\sim 1$ cm and finesse $\sim 4000$ and $\kappa\simeq \kappa_H=(2\pi)1.5$ MHz. We consider an interaction with $^{40}$Ca$^+$ ions on the $3d\; ^3D_{3/2},m_J=+3/2\rightarrow 4p\; ^2P_{1/2},m_J=+1/2$ (probe), $3d\; ^3D_{3/2},m_J=-1/2\rightarrow 4p\; ^2P_{1/2},m_J=+1/2$ (control) and $3d\; ^3D_{3/2},m_J=-1/2\rightarrow 4p\; ^2P_{3/2},m_J=+1/2$ (switching) transitions, for which the respective maximal ion coupling strength are $(g_p,g_c,g_s)=(2\pi)\times(0.53,0.22,0.18)$ MHz respectively.  We assume a standard situation and numerically calculate the steady state normalized probe transmission by solving Eqs.~(\ref{eq:fullfirst})-(\ref{eq:fulllast}) for a probe input field intensity such that the mean intracavity photon number is one in steady state in an empty resonant cavity.

\begin{figure}[h]
\includegraphics[width=\columnwidth]{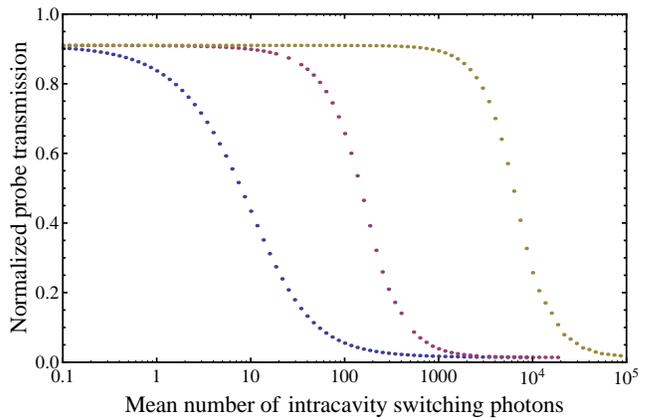}
\caption{(Color online) Probe field normalized transmission as a function of intracavity switching photon number $n_s$ for different switching field detunings $\Delta_s$ (blue: $\Delta_s=(2\pi)0$ MHz, red: $\Delta_s=(2\pi)110$ MHz, green: $\Delta_s=(2\pi)4300$ MHz). Parameters: $(g_p\sqrt{N},\gamma,\gamma_s,\gamma_0,\kappa,\Omega_c,\Delta)=2\pi\times(16,11.2,11,6\times 10^{-4},1.5,2,0)$ MHz.}
\label{fig:switching_numbers}
\end{figure}

Taking an effective collective coupling strength $g\sqrt{N}=(2\pi)16$ MHz renders the crystal/cavity system completely opaque for the probe field in the absence of control field ($\emph{T}_0\sim 1\%$). We assume that the cavity field decay rate is the same for all fields and that the atomic decay rates as defined in (\ref{eq:fullfirst})-(\ref{eq:fulllast}) are $(\gamma,\gamma_s,\gamma_0)=(2\pi)\times(11.2,11,6\times 10^{-4})$ MHz. For the simulations the control and switching fields are injected 0.5 $\mu$s before the probe field, with rise times much shorter than the inverse of the cavity field decay rate, to allow them for reaching their steady state values. The mean intracavity probe photon number is then calculated in steady state, yielding the cavity transmission. Using a control field Rabi frequency $\Omega_c=(2\pi)2$ MHz allows for increasing the resonant probe transmission to $\sim 90 \%$. The variation of the probe transmission for different switching field detunings $\Delta_s$ and intracavity photon numbers $n_s$ is shown in Fig.~\ref{fig:switching_numbers} under these conditions. For all these simulations we checked that the absorption of photons to level $|4\rangle$ was negligible and that the depletion of atoms from level $|1\rangle$ remained at most at the percent level. Optical switching is observed to take place with increasing photon numbers as the detuning is increased, as expected from the previous discussion and analysis.
We define the minimal switching photon number $n_s^*$ as the minimal number of intracavity switching photons needed to bring the normalized transmission from 90 \% to 10 \%. In the \textit{standard} situation, one can easily show from (\ref{eq:chi_EITst}) and (\ref{eq:T_0}) that having a 90 \% transmission in EIT imposes that $\kappa_{EIT}\gtrsim 40\gamma_0$. To get substantial switching we require that the light-shift induced by the switching field $\Omega_s^2/2\Delta_s$ is a few times the width of the EIT transparency window $\kappa_{EIT}$. A numerical estimation shows that $\Omega_s^2/2\Delta_s\sim 5\kappa_{EIT}$, which gives $n_s^*\sim 400\gamma_0\Delta_s/g_s^2$. In agreement with Fig.~\ref{fig:switching_numbers} we find that $\sim 17000$ photons are needed for the large detuning of 4.3 GHz used in~\cite{albert11} and $\sim 400$ for a detuning of $\sim 10\gamma_s$. The previous estimate is actually still valid for a resonant switching field replacing $\Delta_s$ by $\gamma_s$, which would give a minimal photon number of $\sim 40$ for the parameters of Fig.~\ref{fig:switching_numbers}. This illustrative numerical example is based on the experimental parameters of \cite{albert11}, but we note that lower switching numbers could in principle be reached e.g. using smaller cavities or stronger switching transitions

\section{Conclusion}

Using a semiclassical theory for the interaction of four-level atoms with three optical cavity fields, the effect of the transverse mode profiles on the susceptibility and transmission spectrum of a probe field experiencing EIT or EIT-based optical switching has been discussed. Contrarily to the standard situation where the control and switching field Rabi frequencies are the same for all atoms, non-Lorentzian EIT resonance lineshapes and asymmetrical switching lineshapes are predicted when all three fields are coupled to the same cavity mode. Closed analytical forms for the susceptibility and transmission spectrum of the probe field have been found to explain these lineshapes, in good agreement with numerical simulations and with experiments using ion Coulomb crystals in cavities. Last, the prospect for achieving low-light optical switching with ion Coulomb crystals in moderate finesse optical cavities was discussed.
\\

\textit{Acknowledgements} AD is grateful to Thorsten Peters for useful discussions. We acknowledge financial support from the Carlsberg Foundation, the Danish Natural Science Research Council through the European Science Foundation EuroQUAM 'Cavity Mediated Molecular Cooling' project, the European FP7 'Physics of Ion Coulomb Crystals' (PICC) and 'Circuit and Cavity Quantum Electrodynamics' (CCQED) projects.

\appendix
\section{"Localized" or "delocalized" atom situations}\label{sec:appendix}

In this section we investigate the effect of the atomic motion on the cavity EIT feature. Free-space EIT with standing-wave field geometries has been investigated both theoretically~\cite{kocharovskaya01,andre02,zimmer06,moiseev06,hansen07,wu10} and experimentally~\cite{bajcsy03,lin09}, and the influence of the atomic motion on the storage and retrieval of pulses in such geometries has been discussed in e.g.~\cite{zimmer06,hansen07,lin09,wu10}.

\begin{figure}[h]
\includegraphics[width=\columnwidth]{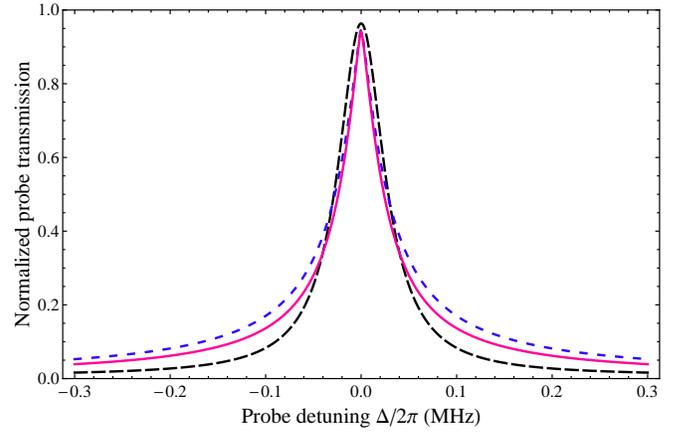}
\caption{(Color online) Probe field transmission spectrum around two-photon resonance for the three situations: \textit{standard} EIT (dotted line) and \textit{all-cavity} EIT with localized (dashed line) and delocalized (solid line) atoms. Parameters as in Fig.~\ref{fig:eit_rabi_trans} with the control field Rabi frequency scaled by a factor 2.2 in the \textit{standard} situation.}
\label{fig:eit_hotcold}
\end{figure}

To discuss the influence of the atoms' longitudinal velocities on the cavity field spectrum in an EIT situation in a relatively simple fashion, one can start from the Hamiltonian (\ref{eq:hamiltonian1}) without switching field and neglecting the variation of the control and probe field Gouy phases and $k$-vector difference ($k_p\simeq k_c=k$)
\begin{align}
\label{eq:hamiltonian1bis}H_{af} =&-\hbar\sum_j g_p\Psi_p(\textbf{r}_j)\cos(kz_j)
\hat{a}_p\hat{\sigma}_{31}^{(j)}\\
\nonumber &-\hbar\sum_j g_c\Psi_c(\textbf{r}_j)\cos(kz_j)\hat{a}_c\hat{\sigma}_{32}^{(j)}+\textrm{h.c.},
\end{align}
Introducing $\hat{\sigma}_{13,\pm}^{(j)}=\hat{\sigma}_{13}^{(j)}\exp(\pm ikz_j)$ and $\hat{\sigma}_{23,\pm}^{(j)}=\hat{\sigma}_{23}^{(j)}\exp(\pm ikz_j)$ the optical dipole operators corresponding to the two running wave fields propagating with $\pm k$, one can recast (\ref{eq:hamiltonian1bis}) into
\begin{align}
\label{eq:hamiltonian1ter}H_{af} =&-\hbar\sum_j g_p\Psi_p(\textbf{r}_j)\hat{a}_p\frac{\hat{\sigma}_{31,+}^{(j)}+\hat{\sigma}_{31,-}^{(j)}}{2}
\\
\nonumber &-\hbar\sum_j g_c\Psi_c(\textbf{r}_j)\hat{a}_c\frac{\hat{\sigma}_{31,+}^{(j)}+\hat{\sigma}_{31,-}^{(j)}}{2}+\textrm{h.c.},
\end{align}
Making the same assumptions as in Sec.~\ref{sec:EIT}, i.e. almost all atoms in state $|1\rangle$, constant and strong control field Rabi frequency, one gets equations of motion in an EIT situation which are similar to Eqs.~(\ref{eq:a_p_EIT},\ref{eq:sigma_13_EIT},\ref{eq:sigma_12_EIT}), but now include the longitudinal dependence of the coupling with the fields:
\begin{align}
\nonumber\dot{a}_p&=&-(\kappa-i\Delta)a_p+ig _p\sum_j\Psi_p(r_j)(\sigma_{13,+}^{(j)}+\sigma_{13,-}^{(j)})/2\\\label{eq:a_p_EITbis}&&+\sqrt{2\kappa_H/\tau}a_p^{in},\\
\nonumber\dot{\sigma}_{13,\pm}^{(j)}&=&-(\gamma-i\Delta)\sigma_{13,\pm}^{(j)}+ig_p\Psi_p(r_j)a_p(1+e^{\pm 2ikz_j})/2\\&&\label{eq:sigma_13_EITbis}+i\Omega_c\Psi_c(r_j)\sigma_{12}^{(j)}(1+e^{\pm 2ikz_j})/2,\\
\label{eq:sigma_12_EITbis}
\dot{\sigma}_{12}^{(j)}&=&-(\gamma_0-i\Delta)\sigma_{12}^{(j)}+i\Omega_c^*\Psi_c(r_j)(\sigma_{13,+}^{(j)}+\sigma_{13,-}^{(j)})/2.\end{align} If the typical timescales for the longitudinal atomic motion (trapping frequencies, thermal motion,...) are faster than the EIT dynamics timescale, but still slower as compared to the atomic dipole dynamics (i.e. if the typical longitudinal velocity $v$ is such that $\kappa_{EIT}\ll|kv|\ll\gamma$), then the terms in $\exp(\pm 2ikz)$ can be averaged out in (\ref{eq:sigma_13_EITbis}) and one retrieves the "delocalized" situation discussed in this paper or in the experiments of~\cite{albert11}. Physically, this can be explained by the fact that the moving atoms will only be in two-photon resonance with the co-propagating parts of the standing waves. As the atoms hence see on average fields with a longitudinal intensity which is half the maximum of the standing wave value, their dipole is reduced which means that the coupling strengths can be effectively rescaled by $1/\sqrt{2}$, yielding the effective Hamiltonian (\ref{eq:hamiltonian2}).

If we now assume that the atoms are sufficiently cold for their longitudinal positions to be fixed with respect to the longitudinal standing-wave structure of the cavity fields during the EIT interaction, one keeps the longitudinal dependance in the coupling terms to solve the previous equations of motion, which are equivalent to
\begin{align}
\nonumber \dot{a}_p&=&-(\kappa-i\Delta)a_p+ig_p\sum_j\Psi_p(r_j)\cos(kz_j)\sigma_{13}^{(j)}\\
\label{eq:a_p_EITcold} &&+\sqrt{2\kappa_H/\tau}a_p^{in},\\
\nonumber\dot{\sigma}_{13}^{(j)}&=&-(\gamma-i\Delta)\sigma_{13}^{(j)}+ig_p\Psi_p(r_j)\cos(kz_j)a_p\\
\label{eq:sigma_13_EITcold} &&+i\Omega_c\cos(kz_j)\Psi_c(r_j)\sigma_{12}^{(j)},\\
\label{eq:sigma_12_EITcold}
\dot{\sigma}_{12}^{(j)}&=&-(\gamma_0-i\Delta)\sigma_{12}^{(j)}+i\Omega_c^*\Psi_c(r_j)\cos(kz_j)\sigma_{13}^{(j)}.
\end{align}
Solving Eqs.~(\ref{eq:a_p_EITcold},\ref{eq:sigma_13_EITcold},\ref{eq:sigma_12_EITcold}) in steady state yields a susceptibility
\begin{equation}
\chi_{EIT}^{cold}=i\sum_j\frac{g_p^2\Psi_p(r_j)^2\cos^2(kz_j)}{\gamma-i\Delta+\frac{\Omega_c^2\Psi_c(r_j)^2\cos^2(kz_j)}{\gamma_0-i\Delta}}
\end{equation}
For a large, uniform density ensemble with random longitudinal ion positions along the cavity axis, one gets
\begin{equation}
\chi_{EIT}=\frac{ig_p^2N}{\gamma-i\Delta}\frac{2\ln[(1+\sqrt{1+2\Theta})/2]}{\Theta}
\end{equation}
where $\Theta$ is given by Eq.~(\ref{eq:Theta}). The probe field transmission spectrum around two-photon resonance is shown in Fig.~\ref{fig:eit_hotcold} for the same parameters as in Fig.~\ref{fig:eit_rabi_trans}b in the three situations considered: \textit{standard} EIT and localized/delocalized \textit{all-cavity} EIT. The localized situation is seen to give rise to a slightly broader EIT resonance, since the atoms see on average a slightly higher effective control field Rabi frequency, as one would intuitively expects.

\end{document}